\documentstyle[12pt]{article}

\makeatletter
\def\section{\@startsection{section}{1}{\z@}{-3.5ex plus -1ex minus -.2ex}
{2.3ex plus .2ex}{\large\bf}}
\makeatother
 
\makeatletter
\@addtoreset{equation}{section}

\makeatother

\def\la{\langle}
\def\ra{\rangle}

\def\lb{\lbrack}
\def\rb{\rbrack}

\def\wh{\widehat}
\def\wt{\widetilde}

 \def\Slash#1{
  \begin{picture}(5,6)(0,0)
  \put(-.7,-1.2){\line(5,6)6}
  \end{picture}
  \kern-.8em#1}
 \def\slash#1{
  \begin{picture}(5,6)(0,0)
  \put(-1.5,-1.7){\line(5,6)5}
  \end{picture}
  \kern-.8em#1}

\def\Sn{\Slash \nabla}

\def\Tr{\mbox{Tr}}
\def\tr{\mbox{tr}}

\def\be{\begin{eqnarray}}
\def\ee{\end{eqnarray}}

\def\g5{\gamma_5}
\def\hg5{\hat{\gamma_5}}

\def\C{{\cal C}}

\def\U{{\cal U}}

\def\D{{\cal D}}

\def\index{\mbox{index}\,}
\def\e{\epsilon}
\def\s{{\cal S}}
\def\B{{\cal B}}
\def\A{{\cal A}}
\def\F{{\cal F}}
\def\N{{\cal N}}
\def\I{{\cal I}}
\def\O{{\cal O}}

\parskip=\medskipamount

\begin{document}

\vspace{8mm}

\begin{center}

{\Large \bf Global obstructions to gauge-invariance in chiral gauge theory
on the lattice}
\\

\vspace{12mm}

{\large David H. Adams}

\vspace{4mm}

Math. dept. and Centre for the Subatomic Structure of Matter, 
University of Adelaide, Adelaide, S.A. 5005, 
Australia. \\

\vspace{1ex}

email: dadams@maths.adelaide.edu.au

\end{center}

\begin{abstract}
It is shown that certain global obstructions to gauge-invariance in chiral
gauge theory, described in the continuum by Alvarez-Gaum\'e and Ginsparg,
are exactly reproduced on the lattice in the Overlap formulation at small
non-zero lattice spacing (i.e. close to the classical continuum limit).
As a consequence, the continuum anomaly cancellation condition $d_R^{abc}=0$ 
is seen to be a necessary (although not necessarily sufficient) condition 
for anomaly cancellation on the lattice in the Overlap formulation.

\end{abstract}

\section{Introduction}

The Overlap formalism \cite{overlap1,overlap2} 
provides a potential solution to
the important problem of constructing chiral gauge theories nonperturbatively
on the lattice. Gauge anomalies are a central issue in this context: 
One would like to show that a gauge-invariant formulation of 
chiral gauge theories on the lattice is possible when the usual (continuum)
anomaly cancellation conditions are satisfied. Conversely, when these 
conditions are not satisfied, one would like to see the continuum anomalies
emerge in the lattice formulation. 
In particular, an interesting test for a lattice formulation of chiral
gauge theory is whether it can capture the global obstructions to 
gauge-invariance of the continuum theory, which reflect the topological
structure of the determinant line bundle over the gauge orbit space
\cite{AG,AS-BF}.

In this paper we show that the overlap reproduces a basic class of such
obstructions, described in the continuum by Alvarez-Gaum\'e and 
Ginsparg \cite{AG}.\footnote{ 
The possibility that this could happen had been previously
mentioned in \cite{overlap2}.}
This is a further demonstration of the ability of the
overlap to reproduce topological features of the continuum 
theory.\footnote{
The parity-invariant overlap formulation of vector gauge theory in odd 
dimensions reproduces the global gauge
anomaly of the continuum theory \cite{Nishimura(parity),Kiku-Neu}, while  
for chiral gauge theory in even dimensions it has been shown to reproduce 
Witten's global anomaly \cite{Neu(Witten),Bar}.}
As a consequence we will see that the usual
(i.e. continuum) local anomaly cancellation condition ((\ref{1.4}) below) 
is a necessary condition for 
local anomaly-free chiral gauge theory on the lattice in the overlap 
formulation.

Global obstructions to the vanishing of local gauge anomalies in the
overlap formulation were previously considered by H. Neuberger in
\cite{Neu(PRD)}. 
An analogue of the the geometric perspective on the continuum chiral 
determinant as a section in a determinant line bundle was described for 
the overlap, and a class of  
global obstructions (which are naturally described in 
this setting) was explicitly constructed for the abelian theory.
These were seen to vanish precisely when the fermion content of the theory
satisfies
\be
\sum_{\alpha}e_{\alpha}^3=0
\label{1.1}
\ee
where the $e_{\alpha}$'s label the irreducible U(1) representations of 
the fermion species. This is precisely the condition for cancellation
of local gauge anomalies in the continuum theory.
Thus, in the abelian case, the continuum anomaly cancellation condition
(\ref{1.1}) is a necessary condition for gauge-invariance of the lattice
chiral determinant in the overlap formulation.\footnote{
Herbert Neuberger has pointed out to me that these obstructions also arise
for nonabelian gauge groups with U(1) subgroups, and that in four dimensions
there are no additional restrictions in the nonabelian case, i.e. if all U(1)
subgroups are free from anomalies then so is the nonabelian group.}
Our result is a nonabelian variant of this.\footnote{ 
Although the class of obstructions that we consider is different: 
the ones in \cite{Neu(PRD)} involve a torus in the orbit space of lattice 
gauge fields whereas ours involve a 2-sphere.}

In the continuum theory, the global obstructions of Alvarez-Gaum\'e and 
Ginsparg arise as follows. Take spacetime to be the Euclidean 4-torus 
$T^4$ (the choice of dimension 4 is for concreteness; everything 
generalises to the $T^{2n}$ case for arbitrary $n$),
gauge group $SU(N)$, and consider a family $\phi_{\theta}$ of gauge 
transformations parameterised by $\theta\in{}S^1$. If the fermion is in
the fundamental representation then each $\phi_{\theta}$ is a map from
$T^4$ to $SU(N)$, and the family of these corresponds to a map
$\Phi:T^5\to{}SU(N)\,$, $\Phi(\theta,x)=\phi_{\theta}(x)$.
The action of $\phi_{\theta}$ on a gauge field $A$ determines a family
$\{A^{\theta}\}_{\theta\in{}S^1}$. The winding number of the phase
of the chiral determinant around this circle-family of gauge fields is an
obstruction to gauge-invariance of the chiral determinant (since if the
determinant is gauge-invariant then it is constant around the family
$\{A^{\theta}\}$ and the winding number vanishes).
In \cite{AG} it was shown that this winding number equals the degree of the 
map $\Phi$. Thus the obstruction is non-vanishing precisely when there
exist maps $\Phi:T^5\to{}SU(N)$ with non-vanishing degree (which happens,
e.g., when $N=3$).\footnote{In \cite{AG} the spacetime was $S^4$ rather
than $T^4\,$, and a condition $\phi_0\equiv1$ was imposed, which allows
$\Phi$ to be viewed as a map from $S^5$ to $SU(N)$. There is no essential
difference with the present case though, since there is an isomorphism
between the homotopy equivalence classes of $\mbox{Map}(S^k,SU(N))$ and
$\mbox{Map}(T^k,SU(N))$.}

In the general case where the fermion content is specified by some 
arbitrary (typically reducible)
representation $R$ of $SU(N)$, the preceding generalises as follows.
Instead of the degree of $\Phi\,$, which is given by an expression of the
form $d^{abc}h_{abc}$ where
\be
d^{abc}=2\tr((T^aT^b+T^bT^a)T^c)
\label{1.2}
\ee
and the $T^a$'s are the generators of $SU(N)$, the obstruction is given by
$d_R^{abc}h_{abc}$ where $d_R^{abc}$ is given by (\ref{1.2}) with
$T^a$ replaced by $R(T^a)$ etc.
Using the well-known fact that there is a relation of the form\footnote{
The existence of a relation of this form can be seen as follows.
Since the representation ring of $SU(N)$ is generated by the fundamental
representation and its complex conjugate, it suffices to show (\ref{1.3})
in the case where $R$ is a tensor product of copies of the fundamental
representation. Then $R(T^a)=T^a\otimes1\otimes\cdots\otimes1+
1\otimes{}T^a\otimes1\otimes\cdots\otimes1+\dots+
1\otimes\cdots\otimes1\otimes{}T^a$ etc, and it follows that
$(R(T^a)R(T^b)+R(T^b)R(T^a))R(T^c)=(T^aT^b+T^bT^a)T^c\otimes1\otimes\cdots
\otimes1+\dots+1\otimes\cdots\otimes1\otimes(T^aT^b+T^bT^a)T^c+$ terms which
have a single $T^a$, $T^b$, or $T^c$ in one of the tensor slots.
Since $\tr(T^a)=0$ etc, it follows that the trace of these latter terms 
vanishes and we get (\ref{1.3}).}
\be
d_R^{abc}=c(R)d^{abc}
\label{1.3}
\ee
we see that the obstruction in the general case is $c(R)$ times the
degree of $\Phi$. Thus in the case where $\mbox{Map}(T^5,SU(N))$ contains
maps with non-vanishing degree a necessary condition for
gauge-invariance of the chiral determinant is $c(R)=0\,$, or
\be
d_R^{abc}=0
\label{1.4}
\ee
Of course, this is just the usual (necessary and sufficient) condition for
anomaly cancellation in the continuum theory (the non-abelian
analogue of (\ref{1.1})).

In this paper we consider a lattice version of the preceding obstructions
in the overlap formulation, and show that they
reduce to the continuum obstructions in the classical continuum limit.
Since the lattice and continuum obstructions are both specified by integers, 
it follows that the lattice obstruction is exactly equal to the continuum one
at small non-zero lattice spacing (i.e. close to the classical continuum
limit). Our approach is similar to the recent analytic work of O. B\"ar and 
I. Campos on the lattice version of Witten's global anomaly \cite{Bar}. 
When combined with the preceding
observation (\ref{1.3}), our result implies that (\ref{1.4}) is a 
necessary condition for gauge-invariance of the lattice chiral determinant
in the overlap formulation, at least in the case where $\mbox{Map}(T^5,SU(N))$
contains maps with non-vanishing degree.\footnote{In higher dimensions there
are cases where the maps all have vanishing degree yet the anomaly 
coefficient $d^{a_1\cdots{}a_n}$ (the symmetrised trace of 
$T^{a_1},\dots,T^{a_n}$) is non-vanishing. E.g. in dimension $2n=6$
with gauge group SU(3) one has $\pi_7SU(3)=0$ and $d^{a_1a_2a_3a_4}\ne0\,$,
cf. p.472 of \cite{AG}.}

It should be emphasised that global obstructions, and hence the results 
of this paper, are
independent of the choice of phase in the overlap chiral determinant.
In contrast, the consistent gauge anomaly for the overlap chiral determinant
does depend on the phase choice. The consistent anomaly in the overlap
formulation, and its classical continuum limit, has been previously studied
in a number of works \cite{Strathdee,Suzuki(rel),Suzuki(recent)}, 
although these have all involved some
form of approximation (e.g. linearisation of the overlap) and/or
assumptions (e.g. weak field, slowly varying field). No such approximations
or assumptions are made in this paper.

The key question which these results lead on to (but which we do not
pursue in this paper) is whether (\ref{1.4}) is a {\it sufficient} 
condition for existence of a local anomaly-free lattice chiral gauge theory 
at non-zero lattice spacing.  
This is currently a topic of major interest and activity 
\cite{Neu(PRD),L1,L2,Suzuki(recent),Neu(recent)}.
To proceed with this in practice,
a specific phase choice must be made to begin
with. (The standard choice is the so-called Brillouin--Wigner 
phase \cite{overlap2}.) 
One can then try to ``improve'' the starting phase choice in various 
ways to get a local anomaly-free overlap when the  
cancellation condition holds. One practical approach is to average along 
the gauge orbits (i.e. the FNN mechanism); see \cite{Neu(PRD)}
and the ref.'s therein.\footnote{
The viability of this approach has been a topic of debate in the literature
\cite{Shamir,Neu(rebutt),Nishimura(trans)}, although there is a body of
evidence which is supportive of it ---see, e.g., \cite{more,Kiku(rebutt)}.}
Another, more theoretical, approach is to reduce the problem 
of going from an arbitrary starting phase choice to 
one for which the (local) anomalies vanish
to that of solving a system of finite-difference equations on the lattice
\cite{L1,L2}.\footnote{
The formulation of ref.'s \cite{L1,L2} (a functional integral formulation 
based on a lattice Dirac operator satisfying the Ginsparg-Wilson relation
\cite{GW}, which had been rediscovered outside of the overlap setting
in the work of Hasenfratz and collaborators \cite{Hasenfratz,Laliena})
is structurally identical to the overlap formulation after identifying the
chiral fermion measures in the functional integral with the many-body
groundstates in the overlap. More on this in \S2, where the many-body
groundstates are the ``unit volume elements'' in our terminology.}
In fact the integrability of these equations has been proved in the
abelian case \cite{L1}.\footnote{
The argument in \cite{L1} relied on a
result on the structure of the abelian axial anomaly \cite{L(axial)},
which has been further elucidated in \cite{Suzuki(Luscher)}.}
There are strong indications that the same 
can be done in the nonabelian case \cite{L2,Suzuki(recent)},
although a complete proof of this has not yet been given. 
Quite recently, a practically-oriented analytic prescription for
constructing anomaly-free non-compact
chiral U(1) gauge theory on the lattice has been given,
starting from a adiabatic phase choice for the overlap \cite{Neu(recent)}.

In \S2 we review the overlap construction of the chiral determinant.
In \S3 the lattice version of the global obstruction
of Alvarez-Gaum\'e and Ginsparg is described, and is shown to reduce to
the continuum obstruction in the classical continuum limit. The derivation
of a key formula used to establish this is given separately in \S4.
This formula ((\ref{a16}) below) is due to L\"uscher \cite{L2},
and our detailed explicit derivation in \S4 is intended to complement
the rather brief argument in \cite{L2}.
In \S5 we make some concluding remarks.
Some details of our calculations are given in an appendix.

\section{Overlap construction of the chiral determinant on the lattice}

The spacetime is taken to be the Euclidean 4-torus $T^4$ 
with fixed edge length $L$.
(Again, the choice of dimension 4 is
for concreteness and simplicity; the arguments and results in the following
generalise straightforwardly to the $T^{2n}$ case for arbitrary $n$.)
We consider hyper-cubic lattices on 
$T^4$ with $2N$ sites along each edge and lattice spacing $a=L/2N$.
\footnote{This $N$ is of course not related to the $N$ in $SU(N)$.}
Given such a lattice, the space of lattice spinor fields $\psi(x)$ is denoted
$\C\,$, and the space of lattice gauge fields $U_{\mu}(x)$ is denoted
$\U$. The space $\C$ is finite-dimensional and comes equipped with an
inner product:
\be
\la\psi_1\,,\psi_2\ra=a^4\sum_x\psi_1(x)^*\psi_2(x)
\label{2.1}
\ee
With suitable boundary conditions, the covariant forward and backward
finite difference operators $\frac{1}{a}\nabla_{\mu}^{\pm}$ act on
$\C$ by
\be
\nabla_{\mu}^+\psi(x)&=&U_{\mu}(x)\psi(x+ae_{\mu})-\psi(x) \label{2.2} \\
\nabla_{\mu}^-\psi(x)&=&\psi(x)-U_{\mu}(x-ae_{\mu})^{-1}\psi(x-ae_{\mu})
\label{2.3} 
\ee
$e_{\mu}$ denotes the unit vector in the positive $\mu$-direction.
We restrict to the case where $U_{\mu}(x)$ and $\psi(x)$ are periodic.
This is the relevant case for considering the classical continuum limit
with topologically trivial gauge fields. Since the chiral determinant
vanishes in the topologically non-trivial case, this suffices for our
purposes. Set $\nabla_{\mu}=\frac{1}{2}(\nabla_{\mu}^++\nabla_{\mu}^-)\,$;
this operator is anti-hermitian with respect to the inner product (\ref{2.1})
since $(\nabla_{\mu}^{\pm})^*=-\nabla_{\mu}^{\mp}$. The Wilson--Dirac 
operator is now given by 
\be
D_{Wilson}=\frac{1}{a}\Sn+\frac{r}{2}a(\frac{1}{a^2}\Delta)
\label{2.4}
\ee
where $\Sn=\sum_{\mu}\gamma^{\mu}\nabla_{\mu}$ (the $\gamma^{\mu}$'s are 
taken to be hermitian so $\Sn$ is anti-hermitian), $\Delta=\sum_{\mu}
\nabla_{\mu}^-+\nabla_{\mu}^+=\sum_{\mu}(\nabla_{\mu}^+)^*\nabla_{\mu}^+=
\sum_{\mu}(\nabla_{\mu}^-)^*\nabla_{\mu}^-$ (hermitian, positive) and
$r>0$ is the Wilson parameter. The hermitian operator
\be
H(m)=\gamma_5(aD_{Wilson}-rm)=\gamma_5(\Sn+r({\textstyle \frac{1}{2}}
\Delta-m))
\label{2.5}
\ee
determines an orthogonal decomposition
\be
\C=\C_+^{(m,U)}\oplus\C_-^{(m,U)}
\label{2.6}
\ee
where $\C_+^{(m,U)}$ and $\C_-^{(m,U)}$ are the subspaces spanned by the
eigenvectors of $H(m)$ with positive and negative eigenvalues respectively.
(We are restricting to the $m,U$ for which $H(m)$ has no zero-modes.)
These subspaces are characterised by $\e(m)=\pm1$ on $\C_{\pm}^{(m,U)}$ where
\be
\e(m)=\frac{H(m)}{\sqrt{H(m)^2}}
\label{2.9}
\ee
(the dependence on $U$ has been suppressed). Noting that
\be
\e(m)=\frac{\frac{1}{|rm|}H(0)-\frac{m}{|m|}\gamma_5}
{\sqrt{(\frac{1}{|rm|}H(0)-\frac{m}{|m|}\gamma_5)^2}}\;\to\;\gamma_5
\qquad\ \mbox{for}\quad{}m\to-\infty 
\label{2.10}
\ee
we see that in the $m\to-\infty$ limit (\ref{2.6}) reduces to the usual
chiral decomposition
\be
\C=\C_+\oplus\C_-
\label{2.10a}
\ee
independent of $U$. Set $m=1$ (the canonical value; $0<m<2$ would suffice)
and let $v_{\pm}$ and $w_{\pm}(U)$ be unit volume 
elements\footnote{A vectorspace $V$
determines vectorspaces $\Lambda^pV$ ($p\!=\!1,\dots,\dim{}V$): the
exterior algebra (=Grassmann algebra) of $V$ of degree $p$. An inner product
in $V$ induces an inner product in each $\Lambda^pV$. A ``unit volume element
on $V$'' is an element $v\in\Lambda^dV$ ($d=\dim{}V$) with $|v|=1$.
E.g. if $v_1,\dots,v_d$ is an orthonormal basis for $V$ then $v_1\wedge\cdots
\wedge{}v_d$ is a unit volume element. Since $\Lambda^dV$ is 1-dimensional,
a unit volume element is unique up to $\pm$ if $V$ is real, or up to a phase
if $V$ is complex.}
on $\C_{\pm}$ and $\C_{\pm}^{(1,U)}$ respectively; these are unique up to 
phase factors. Then the 
lattice versions of the right- and left-handed chiral determinants in
the overlap construction are, respectively,
\be
\la{}v_+\,,w_+(U)\ra&\;&\qquad\qquad\mbox{(right-handed)} \label{2.11} \\
\la{}v_-\,,w_-(U)\ra&\;&\qquad\qquad\mbox{(left-handed)} \label{2.12}
\ee
(see \cite{overlap2} for background and motivation).
The $w_{\pm}(U)$ are required to depend smoothly on $U\,$; then the 
overlaps (\ref{2.11})--(\ref{2.12}) are smooth in $U$. Note that a condition
for non-vanishing overlaps is $\dim\C_{\pm}^{(1,U)}=\dim\C_{\pm}\,\equiv{}\,d$.
The overlaps are unique up to a phase factor, and their norms are 
gauge-invariant (an easy consequence
of the gauge-covariance of $H(m)$, $\e(m)$).

\noindent {\it Remark 2.1.} 
The construction of the overlaps (\ref{2.11})--(\ref{2.12}) requires that
$H(1)$ has no zero-modes. This can be guaranteed by imposing the condition
\cite{L(local),Neu(bound)}
\be
||1-U(p)||\;\le\;0.04\qquad\qquad\forall{}p
\label{2.13}
\ee
on the lattice gauge field $U\,$, where $U(p)$ is the product of the link
variables around a plaquette $p$. This condition 
is automatically satisfied in the classical continuum limit since
$1-U(p_{x;\mu,\nu})=-a^2F_{\mu\nu}(x)+O(a^3)$.

We henceforth restrict $\U$ to be the space of lattice gauge fields
satisfying (\ref{2.13}).

\noindent {\it Remark 2.2.} 
The overlaps (\ref{2.11})--(\ref{2.12}) are determined
(up to a phase) solely by $\e=\e(1)$. The construction could be carried
through given any hermitian operator $\e$ with the property $\e^2=1$.
The norms of the resulting overlaps would be gauge-invariant provided
$\e$ is gauge-covariant.

\noindent {\it Remark 2.3.} 
The overlaps (\ref{2.11})--(\ref{2.12}) can be written 
as\footnote{
We are using the fact that a linear operator $D:W\to{}V$ induces linear
operators $\wh{D}:\Lambda^pW\to\Lambda^pV$ for all $p\,$, defined by
$\wh{D}(w_1\wedge\cdots\wedge{}w_p)=Dw_1\wedge\cdots\wedge{}Dw_p$.
Note that if $W=V$ and $d=\dim{}V$ then $\wh{D}(w_1\wedge\cdots\wedge{}w_d)
=\det{}D\cdot{}w_1\wedge\cdots\wedge{}w_d$.}
\be
({\textstyle \frac{2}{a}})^d
\la{}v_+\,,w_+\ra&=&\la{}v_+\,,\wh{D}w_+\ra
\ \ \equiv\ \det{}D_L \label{2.14} \\
({\textstyle \frac{2}{a}})^d\la{}v_-\,,w_-\ra&=&\la{}v_-\,,\wh{D}w_-\ra
\ \ \equiv\ \det{}D_R \label{2.15} 
\ee
where
\be
D=\frac{1}{a}\Bigl(\,1+\gamma_5\e)
\label{2.16}
\ee
and $\e=\e(1)$ is given by (\ref{2.9}). This follows easily
from the facts that $(1+\gamma_5\e)w=(1\pm\gamma_5)w$ for
$w\in\C_{\pm}^{(1,U)}$ and $(1\pm\gamma_5)v=2v$ for $v\in\C_{\pm}$.
The relations (\ref{2.14})--(\ref{2.15}) show how the overlaps can be
viewed as chiral determinants in an analogous way to the continuum setting:
Set $\hg5=\gamma_5(1-aD)=-\e\,$, then $\hg5^2=1$ and  
$D\hg5=-\gamma_5D\,$, which implies that $D$ maps 
$\wh{C}_{\mp}:=\C_{\pm}^{(1,U)}$ to $\C_{\pm}$. Thus, modulo the factors
$(\frac{2}{a})^d\,$, the right-handed overlap can be viewed as a left-handed
chiral determinant, and vice-versa, as indicated in 
(\ref{2.14})--(\ref{2.15}).\footnote{
A careful consideration of the overlap prescription shows that the overlaps
$\la{}v_{\pm},w_{\pm}\ra$ really should be multiplied by a factor
$(\frac{2}{a})^d$ as in (\ref{2.14})--(\ref{2.15}). These factors are 
physically irrelevant though: they appear both in the numerator and 
denominator in expressions for physical expectation values, and hence
cancel out, and they do not affect anomalies since these only have to
do with the phase of the overlaps. Nevertheless, they are relevant if one
considers the chiral determinant on its own and wishes to use the lattice
regularisation as an alternative to, e.g., zeta-regularisation.}
In this form the overlap arises as the chiral determinant in 
L\"uscher's formulation \cite{L1,L2} after
identifying the unit volume elements $v_{\pm}$ and $w_{\pm}$ 
(the many-body groundstates in the overlap) with the
chiral fermion measures.
These observations have been pointed out previously in \cite{Kiku,Neu(rant)}.
As mentioned there, it is easy to see that an operator $D$ is of the form
(\ref{2.16}), with $\e$ hermitian and $\e^2=1\,$, if and only if it
satisfies the the following two conditions:
\be
D\g5+\g5D&=&aD\g5D\qquad\qquad\ \mbox{(Ginsparg--Wilson relation \cite{GW})}
\label{2.17} \\
D^*&=&\g5D\g5\qquad\qquad\qquad\mbox{($\g5$--hermiticity)}
\label{2.18}
\ee
Also, clearly $D$ is gauge-covariant if and only if $\e$ is gauge-covariant.
The operator
(\ref{2.16}), with $\e=\e(1)$ given by (\ref{2.9}), is the Overlap Dirac
operator introduced by Neuberger in \cite{Neu(PLB)}.
It should also be mentioned that the Ginsparg--Wilson relation was 
rediscovered outside of the overlap setting in the work of P. Hasenfratz
and collaborators ---they considered a different solution, the so-called
perfect Dirac operator \cite{Hasenfratz,Laliena}.

The nullspace of $D$ is invariant under $\g5$ (this follows from the
GW relation (\ref{2.17}): $D\psi\!=\!0\;\Rightarrow\;D(\g5\psi)\!=\!
(aD\g5D-\g5D)\psi\!=\!0$) so 
$\index{}D\equiv\Tr(\g5|_{\ker{}D})$ is well-defined,
as was first noted in \cite{Laliena}. We only need to consider the lattice
gauge fields $U$ for which 
$\dim\C_{\pm}^{(1,U)}=\dim\C_{\pm}\equiv{}d\,$, since
the overlaps vanish otherwise. As noted in \cite{L2}, this 
corresponds to having $\index{}D=0$. Therefore, we henceforth take
$\U$ to be the space of lattice gauge fields satisfying (\ref{2.13}) and 
$\index{}D=0$.

\section{Global obstructions to gauge-invariance of the Overlap}

From now on we consider only the right-handed overlap $\la{}v_+\,,w_+(U)\ra$
(the situation for the left-handed overlap is analogous).
A lattice version of the obstructions considered 
by Alvarez-Gaum\'e and Ginsparg is as follows. Let $\phi_{\theta}$ 
be a family of lattice gauge transformations parameterised by 
$\theta\in{}S^1$. We can assume that the fermion content is specified
by the fundamental representation of $SU(N)$; it will be clear from what
follows that the general case is related to this case in the same way 
as in the continuum setting discussed in the introduction. If $U\in\U$
is a lattice gauge field for which the overlap $\la{}v_+\,,w_+(U)\ra$
is non-vanishing\footnote{
It can be seen from (\ref{2.14}) that $\la{}v_+\,,w_+(U)\ra$ vanishes at the 
$U$ for which $D$ has zero-modes. Generically, these are isolated points
in $\U$.}
then the action of $\phi_{\theta}$ on $U$ determines a map
\be
S^1\to{\bf C}-\{0\}\qquad,\qquad\quad\theta\mapsto
\la{}v_+\,,w_+(\phi_{\theta}\cdot{}U)\ra
\label{a1}
\ee
Since $|\la{}v_+\,,w_+(U)\ra|$ is gauge-invariant, we have
$\la{}v_+\,,w_+(\phi_{\theta}\cdot{}U)\ra=e^{i\alpha(\theta)}
\la{}v_+\,,w_+(U)\ra$ for some phase $\alpha(\theta)\,$, and the map
(\ref{a1}) has integer winding number 
$W(\Phi,U)=\frac{1}{2\pi}(\alpha(1)-\alpha(0))$.
Obviously, if the winding number is non-vanishing $\la{}v_+\,,w_+(U)\ra$
cannot be gauge-invariant. To see that this is a genuine obstruction
to gauge-invariance we note that it is independent of the choice of
$w_+(U)\,$: If $\wt{w}_+(U)$ is another unit volume element on 
$\C_+^{(1,U)}\,$, smoothly varying with $U\,$, then 
$\wt{w}_+(U)=e^{i\beta(U)}w_+(U)$ where the phase factor 
$e^{i\beta(U)}$ is smooth in $U\,$, and we have
\be
\la{}v_+,\wt{w}_+(\phi_{\theta}\cdot{}U)\ra
=e^{i\alpha(\theta)+i\beta(\phi_{\theta}\cdot{}U)}
\la{}v_+\,,w_+(U)\ra
\nonumber
\ee
Assuming that $\{\phi_{\theta}\cdot{}U\}_{\theta\in{}S^1}$ is a contractible
circle in $\U$ (which is certainly true close to the classical continuum 
limit), it follows that this has the same winding number as (\ref{a1})
since since $e^{i\beta(U)}$
is a smooth, non-vanishing, globally defined function of $U$. 
Hence the winding number $W(\Phi,U)$ 
is an obstruction to gauge-invariance of the overlap, independent 
of the choice of $w_+$. 

Our main result is that this obstruction coincides with the continuum one
at small non-zero lattice spacing close to the classical continuum limit:

\vspace{1ex}

\noindent {\it Theorem.} If $\phi_{\theta}$ is the restriction to the 
lattice of a family of continuum gauge transformations (also denoted
$\phi_{\theta}$) and $U$ is the lattice transcript of a topologically
trivial continuum gauge field, then there is an $a_0>0$ (depending on the
$\phi_{\theta}$'s and $U$) such that 
\be
W(\Phi,U)=\deg(\Phi)\quad\quad\mbox{for all}\quad\;a<a_0
\label{a2}
\ee
where $\deg(\Phi)$ is the degree of the continuum map $\Phi:T^5\to{}SU(N)$
given by $\Phi(\theta,x)=\phi_{\theta}(x)$.

\vspace{1ex}

\noindent In light of the discussion in the introduction we conclude
from this that, in the general case where the fermion content is specified
by a general representation $R$ of $SU(N)$, a necessary condition for 
existence of a gauge-invariant construction of the overlap is
\be
d_R^{abc}=0
\label{a3}
\ee

The remainder of the paper is concerned with the proof of the above theorem.
We start by expressing the obstruction as
\be
W(\Phi,U)&=&
\frac{1}{2\pi{}i}\int_0^1d\theta\,{\textstyle \frac{d}{d\theta}}\log\la{}v_+\,,
w_+(\phi_{\theta}\cdot{}U)\ra \nonumber \\
&=&\frac{1}{2\pi{}i}\int_{\s^1}d\,\log\la{}v_+\,,w_+\ra
\label{a4}
\ee
where $\s^1$ denotes the circle $\{\phi_{\theta}\cdot{}U\}_{\theta\in{}S^1}$
in $\U$ and $d$ denotes the exterior derivative on functions (or more 
generally, differential forms) on $\U$. 
After noting that
\be
dw_+=\la{}w_+\,,dw_+\ra{}w_++(dw_+)_{\perp}
\label{a5}
\ee
where $(dw_+)_{\perp}$ denotes the projection of $dw_+$ onto the orthogonal
complement of $w_+$ in $\Lambda^d\C\,$, one finds \cite{Neu(PRD)}
\be
d\,\log\la{}v_+\,,w_+\ra=\frac{\la{}v_+\,,(dw_+)_{\perp}\ra}
{\la{}v_+\,,w_+\ra}+\la{}w_+\,,dw_+\ra
\label{a6}
\ee
The first term on the right-hand side of (\ref{a6}) can be re-expressed as
\be
\frac{\la{}v_+\,,(dw_+)_{\perp}\ra}{\la{}v_+\,,w_+\ra}
=\Tr(dDD^{-1}P_+)
\label{a7}
\ee
where $P_+=\frac{1}{2}(1+\g5)$ is the projection onto $\C_+$. This is a 
straightforward consequence of (\ref{2.14}) and relations noted in
\cite{L1,L2}; for completeness we provide a derivation
in the appendix. Then
\be
d\,\log\la{}w_+\,,w_+\ra=\Tr(dDD^{-1}P_+)+\la{}w_+\,,dw_+\ra\,.
\label{a8}
\ee
Set $w_+^{\theta}=w_+(\phi_{\theta}\cdot{}U)$ and let $D_{\theta}$ denote
$D$ with lattice gauge field $\phi_{\theta}\cdot{}U$. The gauge-covariance
of $D$ gives $\frac{d}{d\theta}D_{\theta}
=\lb\frac{d\phi_{\theta}}{d\theta}\phi_{\theta}^{-1}\,,
D_{\theta}\rb\,$, leading to \cite{L1,L2}
\be
\Tr({\textstyle \frac{d}{d\theta}}D_{\theta}D_{\theta}^{-1}P_+)
&=&\Tr\Bigl(\,\frac{d\phi_{\theta}}{d\theta}\phi_{\theta}^{-1}P_+\Bigr)
-\Tr\Bigl(\,D_{\theta}\frac{d\phi_{\theta}}{d\theta}\phi_{\theta}^{-1}
D_{\theta}^{-1}P_+\Bigr)
\nonumber \\
&=&\Tr\Bigl(\,\frac{d\phi_{\theta}}{d\theta}\phi_{\theta}^{-1}
(P_+-\wh{P}_-)\Bigr)
\nonumber \\
&=&-{\textstyle \frac{1}{2}}a\Tr
\Bigl(\,\frac{d\phi_{\theta}}{d\theta}\phi_{\theta}^{-1}
\g5D_{\theta}\Bigr)
=-{\textstyle \frac{1}{2}}a
\Tr\Bigl(\,\phi_{\theta}^{-1}\frac{d\phi_{\theta}}{d\theta}
\g5D_1\Bigr)
\label{a9}
\ee
where we have used the fact that $P_+D=D\wh{P}_-$ where 
$\wh{P}_-=\frac{1}{2}(1-\hg5)\,$, $\hg5=\g5(1-aD)$. Substituting (\ref{a8})
into (\ref{a4}) and using (\ref{a9}) we get
\be
2\pi{}iW(\Phi,U)&=&\int_{\s^1}\Tr(dDD^{-1}P_+)+\la{}w_+\,,dw_+\ra
\label{a10} \\
&=&\int_0^1d\theta\,\Bigl(\,-{\textstyle \frac{1}{2}}a
\Tr\Bigl(\,\phi_{\theta}^{-1}\frac{d\phi_{\theta}}{d\theta}
\g5D_1\Bigr)+\la{}w_+^{\theta}\,,\frac{dw_+^{\theta}}{d\theta}\ra\,\Bigr)
\label{a11}
\ee
We have derived this relation under the assumption that the overlap is 
non-vanishing for $U\,$, or equivalently, that the Overlap Dirac operator
$D$ with lattice gauge field $U$ has no zero-modes. By construction
$W(\Phi,U)$ is clearly smooth, and therefore locally constant, in such $U$.
But it is ill-defined at the (generically isolated) points in $\U$ where
the overlap vanishes. One such point is the trivial field $U=1$ (in this
case the zero-momentum spinors with definite chirality are zero-modes
for $D$). However, the right-hand side of (\ref{a11}) is clearly smooth
in $U$ for all $U\in\U$ (since $D$ is smooth in the lattice gauge field when
(\ref{2.13}) is satisfied \cite{L(local)}), and must therefore be a locally
constant function of $U$ for all $U\in\U$. In the continuum, any 
topologically trivial gauge field can be continuously deformed to the 
trivial field. It follows that when the lattice spacing $a$ is sufficiently 
small, the lattice transcript $U$ can be continuously deformed to the
trivial lattice gauge field (using the lattice transcript of the continuum 
path). Therefore, to prove the theorem it suffices to show
that there is an $a_0>0$ such that 
\be
W(\Phi)=\deg(\Phi)\qquad\quad\mbox{for all}\quad\;a<a_0
\label{a11.5}
\ee
where $W(\Phi)$ denotes
the right-hand side of (\ref{a11}) with trivial field $U=1$.
In this case $D_1$ acts trivially in colour space, and 
since $\phi_{\theta}^{-1}\frac{d\phi_{\theta}}{d\theta}$ is
in the Lie algebra of $SU(N)$ the trace over colour indices
in the first term in (\ref{a11}) vanishes, resulting in
\be
2\pi{}iW(\Phi)
&=&\int_0^1d\theta\,\la{}w_+^{\theta}\,,{\textstyle \frac{d}{d\theta}}
w_+^{\theta}\ra \nonumber \\
&=&\int_{\s^1}\la{}w_+\,,dw_+\ra
\label{a12}
\ee
A calculation gives
\be
d\la{}w_+\,,dw_+\ra&=&\la{}dw_+\,,dw_+\ra \nonumber \\
&=&\Tr(PdPdP)
\label{a13}
\ee
where
\be
P\;\equiv\;\wh{P}_-={\textstyle \frac{1}{2}}(1-\hg5)
={\textstyle \frac{1}{2}}(1+\e)
\label{a14}
\ee
with $\e=\e(1)$ given by (\ref{2.9}). The last equality in (\ref{a13})
is derived in the appendix. A simpler version of it (originating in
\cite{Simon}) was used in \cite{Neu(PRD)}; the same relation in a 
different guise was subsequently noted in \cite{L1,L2},
and in more detail in \cite{Suzuki(rel)}.
Now, by Stokes theorem, if $\B^2$ is a disc in $\U$ with boundary $\s^1$
it follows that
\be
2\pi{}iW(\Phi)&=&\int_{\B^2}\Tr(PdPdP) \nonumber \\
&=&\int_{\B^2}\Tr(P\lb\partial_{\theta}P,\partial_tP\rb)\,d\theta{}dt
\label{a15}
\ee
where $(\theta,t)$ are taken to be polar coordinates on a unit disc $B^2$
parameterising $\B^2$.
We take $\B^2$ to be the lattice transcript of a disc-family of continuum
gauge fields, also denoted $\B^2\,$, given by
\be
A^{(\theta,t)}=f(t)\phi_{\theta}d_x\phi_{\theta}^{-1}
\qquad\quad(\theta,t)\in{}B^2
\label{a17}
\ee
where $f(t)$ is an arbitrary smooth function equal to 1 in a
neighbourhood of $t\!=\!1$ and vanishing
in a neighbourhood of $t\!=\!0$. The lattice transcript 
$U^{(\theta,t)}$ has the property $U^{(\theta,1)}=\phi_{\theta}\cdot1$
so the boundary of $\B^2$ is $\s^1$ as required in (\ref{a15}).
Note that (\ref{a15}) is manifestly independent of the choice of $w_+(U)\,$,
i.e. independent of the choice of phase in the overlap.

A general formula for the classical continuum limit of the integrand in
(\ref{a15}) has been given by L\"uscher in \cite{L2}:
If $U^{(s,t)}$ is the lattice transcript of a family $A^{(s,t)}$
of continuum gauge fields, and $P=P^{(s,t)}$ is the corresponding family
of projection operators (given by (\ref{a14})), then
\be
\lim_{a\to0}\;\Tr(P\lb\partial_sP,\partial_tP\rb)
=\frac{-1}{32\pi^2}\int_{T^4}d^4x\,\epsilon_{\mu\nu\rho\sigma}
d^{abc}\partial_sA_{\mu}^a(x)\partial_tA_{\nu}^b(x)F_{\rho\sigma}^c(x)\,.
\label{a16}
\ee
Using this, (\ref{a11.5}) follows easily from (\ref{a15}) as we will see below,
thereby proving the theorem. 
In \cite{L2} the locality, smoothness and symmetry properties
of $P$ were used to show that the limit in (\ref{a16}) exists and is
given by the integral over $T^4$ of a polynomial in the gauge fields and 
its derivatives. However, the explicit form of this polynomial (i.e. the
integrand on the right-hand side of (\ref{a16})) was not obvious, at least
to the present author, from the brief argument in \cite{L2}.
Since (\ref{a16}) is an important formula in this context (it was also 
a key ingredient in the arguments of ref.'s \cite{L2} and
\cite{Bar}) we will give a detailed, explicit derivation of it in \S4.
This is intended to complement the brief argument in \cite{L2}.

By (\ref{a16}), the classical continuum limit of (\ref{a15}) is
\be
\lim_{a\to0}\ 2\pi{}iW(\Phi)=
\frac{-1}{32\pi^2}\int_{\B^2\times{}T^4}d\theta{}dtd^4x\,
\epsilon_{\mu\nu\rho\sigma}d^{abc}\partial_{\theta}
A_{\mu}^a(x)\partial_tA_{\nu}^b(x)F_{\rho\sigma}^c(x)\,.
\label{a18}
\ee
with $A=A^{(\theta,t)}$ given by (\ref{a17}). It remains to show that this
is equal to $2\pi{}i\deg(\Phi)$. 
Then, since $W(\Phi)$ is integer, we can conclude 
that $W(\Phi)=\deg(\Phi)$ for all lattice spacings $a$ smaller than some
$a_0>0$ and the theorem is proved. The right-hand side of (\ref{a16}) can
be shown to equal $2\pi{}i\deg(\Phi)$ by a direct calculation \cite{Bar},
but it is easier and perhaps more illuminating to proceed
indirectly as follows.
We view the family $A^{(\theta,t)}$ as a gauge field on $B^2\times{}T^4\,$:
\be
\A(x,\theta,t)=A_{\mu}^{(\theta,t)}(x)dx^{\mu}
=f(t)\phi_{\theta}(x)\partial_{\mu}\phi_{\theta}^{-1}(x)dx^{\mu}
\label{a19}
\ee
and define another gauge field on $\wt{B}^2\times{}T^4$ by 
\be
\wt{\A}(x,\theta,s)=-f(s)\partial_{\theta}\phi_{\theta}^{-1}(x)
\phi_{\theta}(x)d\theta
\label{a20}
\ee
where $\wt{B}^2$ is another copy of the unit disc,
with polar coordinates $(\theta,s)$. 
$B^2\times{}T^4$ and $\wt{B}^2\times{}T^4$ can be glued together along
their common boundary $S^1\times{}T^4$ to get the closed manifold
$S^2\times{}T^4$. On the common boundary $S^1\times{}T^4$
the fields $\A$ and $\wt{\A}$ are related by a gauge transformation:
\be
\wt{\A}=\Phi^{-1}\cdot\A=\Phi^{-1}\A\Phi+\Phi^{-1}(d_{\theta}+d_x)\Phi
\nonumber
\ee
where $\Phi:\s^1\times{}T^4\to{}SU(N)$ is given by
$\Phi(\theta,x)=\phi_{\theta}(x)$. Therefore $\A$ and $\wt{\A}$ constitute
a gauge field $\wh{\A}$ on an $SU(N)$ bundle over $S^2\times{}T^4$ with
topological charge $-\deg(\Phi)$. The topological charge is also given
by the integral of the Chern character over $S^2\times{}T^4\,$, thus
\be
-\deg(\Phi)&=&\frac{i^3}{(2\pi)^33!}\int_{S^2\times{}T^4}
\tr\,\wh{\F}^3 \nonumber \\
&=&\frac{i^3}{(2\pi)^33!}
\bigg\lb\,\int_{B^2\times{}T^4}\tr{}\F^3\ -\ \int_{\wt{B}^2\times{}T^4}
\tr\,\wt{\F}^3\,\bigg\rb
\label{a21}
\ee
The second term vanishes: $\wt{\F}^3=0$ since $\wt{\A}$ only involves the
1-form $d\theta$. Regarding the first term, from
$\F=(d_x+d_{\theta}+d_t)\A+\A\wedge\A$ we get
\be
\tr\,\F^3&=&\tr(d_{\theta}Ad_tAF+d_tAd_{\theta}AF+d_{\theta}AFd_tA
+d_tAFd_{\theta}A+Fd_{\theta}Ad_tA+Fd_tAd_{\theta}A) \nonumber \\
&=&3\Bigl(\frac{d^{abc}}{2}\Bigr)d_{\theta}A^ad_tA^bF^c \nonumber
\ee
After substituting this for $\tr\,\F^3$ in (\ref{a21}) we see from
(\ref{a18}) that $\lim_{a\to0}W(\Phi)=\deg(\Phi)$ as required.

\section{A detailed derivation of L\"uscher's formula}

In this section we give a detailed, explicit derivation of the formula 
(\ref{a16}). We begin by noting that
\be
\Tr(P\lb\partial_sP,\partial_tP\rb)
&=&\frac{1}{8}\Tr(\e\lb\partial_s\e,\partial_t\e\rb)
\nonumber \\
&=&\frac{1}{4}\Tr(\e\partial_s\e\partial_t\e)
\label{x0}
\ee
Setting $H=H(1)$ we have 
\be
\e=\frac{H}{\sqrt{H^2}}=\frac{\g5{}X}{\sqrt{X^*X}}
\quad\quad,\qquad\ X=a\g5{}H=\Sn+r({\textstyle \frac{1}{2}}\Delta-1)
\label{x0.5}
\ee
and a calculation using (\ref{x0}) gives
\be
&&\Tr(P\lb\partial_sP,\partial_tP\rb) \nonumber \\
&&\quad\ =\frac{1}{4}\Tr\Bigl(\,\frac{\g5{}X}{\sqrt{X^*X}}
\partial_s(\g5{}X)\frac{1}{\sqrt{X^*X}}\partial_t(\g5{}X)\frac{1}{\sqrt{X^*X}}
\Bigr) \label{x3} \\
&&\qquad\ +\frac{1}{4}\Tr\Bigl(\,\partial_s\Bigl(\,\frac{1}{\sqrt{X^*X}}\Bigr)
\partial_t(\g5{}X)\Bigr)
+\frac{1}{4}\Tr\Bigl(\,\frac{1}{X^*X}\partial_s(\g5{}X)\partial_t
\sqrt{X^*X}\Bigr) \label{x4} \\
&&\qquad\ -\frac{1}{4}\Tr\Bigl(\,\partial_s\Bigl(\,\frac{1}{\sqrt{X^*X}}\Bigr)
\sqrt{X^*X}\g5{}X\partial_t\Bigl(\,\frac{1}{\sqrt{X^*X}}\Bigr)\Bigr)
\label{x5}
\ee
We will see below that $(\ref{x4})\sim{}O(a)$ and 
$(\ref{x5})=Symm+O(a)$ where $Symm$ is symmetric under interchange
of $\partial_s$ and $\partial_t$. Since 
$\Tr(P\lb\partial_sP,\partial_tP\rb)$ is antisymmetric under this
interchange it follows that
\be
\Tr(P\lb\partial_sP,\partial_tP\rb)
&=&\frac{1}{8}\bigg\lb
\Tr\Bigl(\,\frac{\g5{}X}{\sqrt{X^*X}}
\partial_s(\g5{}X)\frac{1}{\sqrt{X^*X}}\partial_t(\g5{}X)\frac{1}{\sqrt{X^*X}}
\Bigr) \nonumber \\
&&-\Tr\Bigl(\,\frac{\g5{}X}{\sqrt{X^*X}}
\partial_t(\g5{}X)\frac{1}{\sqrt{X^*X}}\partial_s(\g5{}X)\frac{1}{\sqrt{X^*X}}
\Bigr)\bigg\rb+O(a)
\label{x6}
\ee
To prove these statements, and evaluate the $a\to0$ limit of (\ref{x6}),
we use the fact that the lattice transcript\footnote{
Here $A=A^{(s,t)}$ and $U=U^{(s,t)}$ depend smoothly on the two parameters 
$(s,t)$.}
\be
U_{\mu}(x)=T\,\exp\Bigl(\,\int_0^1aA_{\mu}(x+(1-\tau)ae_{\mu})\,d\tau\Bigr)
\label{x1}
\ee
can be expanded in powers of $a$ as
\be
U_{\mu}(x)=\sum_{n=0}^{\infty}a^n\int_{0<\tau_1<\cdots<\tau_n<1}
d\tau_n\cdots{}d\tau_1\,A_{\mu}(x,\tau_n)\cdots{}A_{\mu}(x,\tau_1)
\label{x7}
\ee
where $A_{\mu}(x,\tau)=A_{\mu}(x+(1-\tau)ae_{\mu})$. Since $A_{\mu}(x)$ is a 
smooth function on the closed manifold $T^4$ there is a finite $K$
such that $|A_{\mu}(x)|<K$ for all $x,\mu$. Then the norm of the integral
in the $n$'th term of (\ref{x7}) is bounded by $\frac{1}{n!}K^n\,$, so
(\ref{x7}) is norm-convergent for all $a$. 
The inverse $U_{\mu}(x)^{-1}$ also has an expansion in powers of $a\,$:
using the fact that $U_{\mu}(x)^{-1}$ is the parallel transport from 
$x$ to $x+ae_{\mu}$ specified by $A$ we see that $U_{\mu}(x)^{-1}$
is given by the right-hand side of (\ref{x7}) with 
$A_{\mu}(x,\tau)=A_{\mu}(x+\tau{}ae_{\mu})$.

Substituting these expansions in (\ref{2.2})--(\ref{2.3}) gives an expansion
$\nabla_{\mu}^{\pm}=\sum_{n=0}^{\infty}a^n(\nabla_{\mu}^{\pm})_n$.
This in turn gives an expansion $X=\sum_{n=0}^{\infty}a^nX_n$.
It is not difficult to show that the $||(\nabla_{\mu}^{\pm})_n||$'s
and the $||X_n||$'s have a finite bound $K'$ independent of $a$ and $n$,
so the expansions are norm-convergent when $a$ is sufficiently small.

To expand $1/\sqrt{X^*X}$ we note that 
\be
X^*X=L+V
\label{x8}
\ee
where
\be
L&=&-\nabla^2+r^2({\textstyle \frac{1}{2}}\Delta-m)^2 
\label{x9} \\
V&=&V^{(1)}+V^{(2)}
\label{x10} \\
V^{(1)}&=&{\textstyle \frac{1}{2}}r\gamma^{\mu}\lb\nabla_{\mu}\,,
\sum_{\nu}\nabla_{\nu}^+-\nabla_{\nu}^-\rb\quad,\quad{}
V^{(2)}=-{\textstyle \frac{1}{4}}\lb\gamma^{\mu}\,,\gamma^{\nu}\rb\,
\lb\nabla_{\mu}\,,\nabla_{\nu}\rb
\label{x11}
\ee
Just as for $X\,$, the expansion (\ref{x7}) leads to expansions
$X^*X=\sum_{n=0}^{\infty}a^n(X^*X)_n\,$, $L=\sum_{n=0}^{\infty}a^nL_n\,$,
$V=\sum_{n=0}^{\infty}a^nV_n$ where the $||(X^*X)_n||$'s, $||L_n||$'s
and $||V_n||$'s again have a finite bound independent of $a$ and $n$.
Furthermore, explicit calculations show that
\be
\lb\nabla_{\mu}^+\,,\nabla_{\nu}^+\rb\psi(x)
&=&(-a^2F_{\mu\nu}(x)+O(a^3))\psi(x+ae_{\mu}+ae_{\nu})
\label{q7} \\
\lb\nabla_{\mu}^+\,,\nabla_{\nu}^-\rb\psi(x)
&=&(-a^2F_{\mu\nu}(x)+O(a^3))\psi(x+ae_{\mu}-ae_{\nu})
\label{q8} \\
\lb\nabla_{\mu}^-\,,\nabla_{\nu}^+\rb\psi(x)
&=&(-a^2F_{\mu\nu}(x)+O(a^3))\psi(x-ae_{\mu}+ae_{\nu})
\label{q9} \\
\lb\nabla_{\mu}^-\,,\nabla_{\nu}^-\rb\psi(x)
&=&(-a^2F_{\mu\nu}(x)+O(a^3))\psi(x-ae_{\mu}-ae_{\nu})
\label{q10} 
\ee
It follows that $V_0=V_1=0\,$, i.e. the expansion of $V$ starts with the
$a^2$ term, hence $||V||\sim{}O(a^2)$. The leading term $a^2V_2$ 
is explicitly given (mod $O(a^3)$) by substituting
(\ref{q7})--(\ref{q10}) into (\ref{x10})--(\ref{x11}). We note from this
that $V_2=V_2^bT^b$ where the $T^b$'s are the generators of the Lie algebra
of $SU(N)$ and the $V_2^b$'s are trivial in colour space.

From (\ref{x7}) we also get expansions of $\partial_sU_{\mu}(x)$
and $\partial_sU_{\mu}(x)^{-1}=U_{\mu}(x)^{-1}\partial_sU_{\mu}(x)
U_{\mu}(x)^{-1}$
in powers of $a\,$, leading to an expansion $\partial_sX=\sum_{n=0}^{\infty}
a^n(\partial_sX)_n$ and expansions of $\partial_s(X^*X)\,$,
$\partial_sL\,$, and $\partial_sV$. Note that these begin with the order
$a$ term, i.e. $(\partial_sX)_0=(\partial_s(X^*X))_0=(\partial_sL)_0
=(\partial_sV)_0=0$.
For later use we also note the following: (1) The lowest order term in the 
expansion of $\partial_sU_{\mu}(x)$ (or $\partial_sU_{\mu}(x)^{-1}$) is 
$aA_{\mu}^b(x)T^b$ (or $-aA_{\mu}^b(x)T^b$). (2) Applying $\partial_s$
to (\ref{q7})--(\ref{q10}) results in $F_{\mu\nu}(x)\to
\partial_sF_{\mu\nu}(x)\,$, so $(\partial_sV)_0=(\partial_sV)_1=0\,$,
$||\partial_sV||\sim{}O(a^2)\,$, and $(\partial_sV)_2=
(\partial_sV)_2^bT^b$ where the $(\partial_sV)_2^b$'s are trivial in colour
space.

Note that the $\gamma^{\mu}$'s in (\ref{x8}) are all contained in $V$.
The hermitian positive operator $L$ is trivial in Dirac indices and the
lowest order term $L_0$ in its expansion is diagonal with respect to
the plane wave basis $\{e^{iak\cdot{}x}\}\,$; the diagonal elements are
\be
L_0(ak)=\sum_{\mu}\sin^2(ak_{\mu})+r^2\Bigl(\,-1+\sum_{\mu}1-\cos(ak_{\mu})
\Bigr)^2
\label{q11}
\ee
From this we see that there is a $b>0$ independent of
$a$ such that $L_0>2b$. Then, by taking $a$ to be sufficiently small,
we can achieve $L>b$ and $||V||<\frac{1}{2}b\,$, in which case
$1/\sqrt{X^*X}$ can be expanded as follows:
\be
\frac{1}{\sqrt{X^*X}}
&=&\int_{-\infty}^{\infty}\frac{d\sigma}{\pi}\,\frac{1}{X^*X+\sigma^2}
\nonumber \\
&=&\int_{-\infty}^{\infty}\frac{d\sigma}{\pi}
\Bigl(\,\frac{1}{1+(L+\sigma^2)^{-1}V}\Bigr)\Bigl(\,\frac{1}{L+\sigma^2}
\Bigr)
\nonumber \\
&=&\int_{-\infty}^{\infty}\frac{d\sigma}{\pi}\,
\sum_{k=0}^{\infty}(-1)^k((L+\sigma^2)^{-1}V)^k\,\frac{1}{L+\sigma^2}\,.
\label{q12}
\ee
For all $p$ we have
\be
\Big|\Big|\sum_{k=0}^p(-1)^k((L+\sigma^2)^{-1}V)^k\,\frac{1}{L+\sigma^2}
\Big|\Big|\;<\;\sum_{k=0}^{\infty}(\frac{1}{b}
({\textstyle \frac{1}{2}}b))^k\,\frac{1}{b+\sigma^2}
=\frac{2}{b+\sigma^2}
\nonumber 
\ee
Since the integral of this over $(-\infty,\infty)$ is finite, the
integral and sum in (\ref{q12}) can be interchanged, resulting in a 
norm-convergent expansion in powers of $V\,$:
\be
(X^*X)^{-1/2}=\sum_{k=0}^{\infty}\int_{-\infty}^{\infty}
\frac{d\sigma}{\pi}\,(-1)^k((L+\sigma^2)^{-1}V)^k(L+\sigma^2)^{-1}
\label{q14}
\ee
Since $||V||\sim{}O(a^2)$ the $k$'th term in the sum is $\sim{}O(a^{2k})$
and we conclude that
\be
(X^*X)^{-1/2}=\sum_{k=0}^p\int_{-\infty}^{\infty}\frac{d\sigma}{\pi}
(-1)^k((L+\sigma^2)^{-1}V)^k(L+\sigma^2)^{-1}+R_{p+1}
\label{q15}
\ee
where $\frac{1}{a^{2p}}||R_{p+1}||\to0$ for $a\to0$.
Similarly, we find
\be
\partial_s(X^*X)^{-1/2}
=\sum_{k=0}^p\int_{-\infty}^{\infty}\frac{d\sigma}{\pi}
(-1)^k\partial_s\Bigl(\,((L+\sigma^2)^{-1}V)^k(L+\sigma^2)^{-1}\Bigr)
+\partial_sR_{p+1}
\label{q15a}
\ee
where $\frac{1}{a^{2p}}||\partial_sR_{p+1}||\to0$ for $a\to0$.

The bound $||\nabla_{\mu}^{\pm}||\le2$ and triangle inequalities lead to
an $a-$independent upper bound $L<b_1$. Using this, the operator 
$(L+\sigma^2)^{-1}$ in (\ref{q15}) can be expanded as
\be
\frac{1}{L+\sigma^2}
&=&\Bigl(\,\frac{1}{b_1+\sigma^2}\Bigr)\Bigl
(\,\frac{1}{1-\frac{b_1-L}{b_1+\sigma^2}}\Bigr) \nonumber \\
&=&(b_1+\sigma^2)^{-1}\sum_{m=0}^{\infty}(b_1+\sigma^2)^{-m}(b_1-L)^m
\label{q16}
\ee
Substituting the expansion $L=\sum_{n=0}^{\infty}a^nL_n$ in (\ref{q16}),
and then substituting in (\ref{q15}) the resulting expansion of 
$(L+\sigma^2)^{-1}\,$, along with the expansion $V=\sum_{n=2}^{\infty}
a^nV_n\,$, we get an expansion $(X^*X)^{-1/2}=\sum_{n=0}^{\infty}
a^n(X^*X)_n^{-1/2}$. Similarly, after applying $\partial_s$ to
(\ref{q16}) and substituting the resulting expansion in (\ref{q15a}),
we get an expansion $\partial_s(X^*X)^{-1/2}=\sum_{n=1}^{\infty}
a^n(\partial_s(X^*X)^{-1/2})_n$ (note $(\partial_s(X^*X)^{-1/2})_0=0$).
These, together with the expansions of $X$ and $\partial{}X$, lead to
expansions $\O=\sum_{n=0}^{\infty}a^n\O_n$ of the operators in
(\ref{x3})--(\ref{x5}). It can be shown that there is a finite $\wt{K}$
independent of $a$ and $n$ such that $||\O_n||<\wt{K}$ for these operators.
This technical result will be presented elsewhere \cite{prep}\footnote{
It relies on the fact that $A_{\mu}(x)$ is periodic. The general (i.e.
topologically non-trivial) case is more complicated.}.
This implies that the aforementioned operator expansions are all 
norm-convergent when $a$ is sufficiently small. An immediate consequence
is the following: In the resulting expressions $\sum_{n=0}^{\infty}
a^n\Tr(\O_n)$ for (\ref{x3})--(\ref{x5}) the part $\sum_{n=5}^{\infty}
a^n\Tr(\O_n)$ vanishes in the $a\to0$ limit. To see this, let 
$\{\psi_j\}_{j=1,\dots,\N}$ be an arbitrary orthonormal basis for $\C\,$;
then
\be
\Big|\sum_{n\ge5}a^n\Tr(\O_n)\Big|
&\le&\sum_{n\ge5,\,j}a^n|\la\psi_j\,,\O_n\psi_j\ra| \nonumber \\
&\le&\sum_{n\ge5,\,j}a^n\wt{K}=a^4\N\wt{K}\sum_{n=0}^{\infty}a^n
=4L^4\wt{K}\frac{a}{1-a}\to0\quad\mbox{for}\ a\to0 \nonumber \\
&&\label{q17}
\ee
where we have used the fact that 
$\N\equiv\dim\C=4(L/a)^4$. This shows that we
only need to consider the $a^n$ terms with $n\le4$ in the expansions
$\sum_{n=0}^{\infty}a^n\Tr(\O_n)$ of (\ref{x3})--(\ref{x5}).

To proceed with the determination of (\ref{x3})--(\ref{x5}) in the
$a\to0$ limit, we note the following: (i) Due to the presence of an odd
number of $\g5$'s, terms in the expansions involving a product of less
that 4 $\gamma^{\mu}$'s vanish. (ii) $L\,$, $V^{(1)}$ and $V^{(2)}$ are
are of order 0, 1 and 2 respectively in the $\gamma^{\mu}$-matrices, c.f.
(\ref{x11}). (iii) In the expansion $\O=\sum_{n=0}^{\infty}a^n\O_n$
of any operator $\O$ constructed from the $\nabla_{\mu}^{\pm}$'s, 
the term $\O_0$ is independent of the gauge field, so $(\partial_s\O)_0=0$.
Hence non-vanishing terms in such expansions are at least $O(a)$ for $a\to0$.
(iv) As we have seen in (\ref{q17}),
terms in the operator expansions which are of order $\ge5$ in $a$ give
vanishing contributions in the $a\to0$ limit.

At this point we can derive the postulated formula (\ref{x6}). Consider the
first trace in (\ref{x4}): After substituting the expansion (\ref{q14})
for $(X^*X)^{-1/2}$ only the terms with at least two $V$'s are non-vanishing 
after taking the trace over spinor indices. Since $||V||\sim{}O(a^2)\,$,
such terms are all of order $\ge4$ in $a$. Since 
$\partial_t(\g5{}X)\,\sim\,O(a)$ it follows that the terms in the expansion
of $\partial_s(X^*X)^{-1/2}\partial_t(\g5{}X)$ which are non-vanishing 
after taking the trace are all of order $\ge5$ in the $a\to0$ limit, so
(\ref{x4})$\sim{}O(a)$ as claimed.

We now consider the trace (\ref{x5}). First, note from (\ref{q15})--(\ref{q16})
that the lowest order term involving $V$ (or containing $\gamma^{\mu}$'s)
in the expansion of $(X^*X)^{-1/2}$ is 
\be
-\int_{-\infty}^{\infty}\frac{d\sigma}{\pi}\,
(L_0+\sigma^2)^{-1}a^2V_2(L_0+\sigma^2)^{-1}\,.
\label{q17.5}
\ee
Using (\ref{q16}) and the fact that $\lb{}L_0,V\rb\sim{}O(a)$ we find
that, modulo an $O(a)$ term, this is
\be
-a^2\int_{-\infty}^{\infty}\frac{d\sigma}{\pi}\,
(L_0+\sigma^2)^{-2}V_2
=-a^2L_0^{-3/2}V_2\int_{-\infty}^{\infty}
\frac{d\sigma}{\pi}\,\frac{1}{(1+\sigma^2)^2}
=-\frac{1}{2}a^2L_0^{-3/2}V_2
\label{q18}
\ee
Similarly, the lowest order term containing $\gamma^{\mu}$'s in the expansion
$\partial_s(X^*X)^{-1/2}$ is $-\frac{1}{2}a^2L_0^{-3/2}(\partial_sV)_2$.
To simplify the notation in the following, we write $\partial_s\O_n$
for $(\partial_s\O)_n$. Now, using (i)--(iv) above, we find
\be
&&\Tr\Bigl(\,\partial_s(X^*X)^{-1/2}(X^*X)^{1/2}\g5{}X\partial_t(X^*X)^{-1/2}
\Bigr) \nonumber \\
&&\ \ =a^4\Tr\Bigl(\,\partial_s(X^*X)_2^{-1/2}(X^*X)_0^{1/2}\g5{}X_0
\partial_t(X^*X)_2^{-1/2}\Bigr)+O(a) \nonumber \\
&&\ \ =\frac{a^4}{4}\Tr\Bigl(\,L_0^{-3/2}\partial_sV_2L_0^{1/2}\g5{}X_0
L_0^{-3/2}\partial_tV_2\Bigr)+O(a)
\label{q19}
\ee
where we have also used $(X^*X)_0=L_0$.
We now supplement (i)--(iv) above with the following observation (v):
$V_2$ and $(\partial{}V)_2$ commute with $L_0$ modulo an $O(a)$ term,
and commute with $\g5{}X_0$ modulo an $O(a)$ term and a term of order
1 in the $\gamma^{\mu}$'s. It follows that, modulo an $O(a)$ term, 
(\ref{q19}) is symmetric under interchange of $\partial_s$ and $\partial_t$
as claimed. This proves the previously stated symmetry property of 
(\ref{x5}), thereby completing the derivation of (\ref{x6}).

Turning now to the traces in (\ref{x6}), similar arguments to the preceding
give
\be 
&&\Tr\Bigl(\,\g5{}X\,(X^*X)^{-1/2}\,\partial_s(\g5{}X)\,(X^*X)^{-1/2}\,
\partial_t(\g5{}X)\,(X^*X)^{-1/2}\Bigr)
\nonumber \\
&&\ =a^4\Tr\Bigl(\,\g5X_0\,(X^*X)_2^{-1/2}\,\partial_s(\g5{}X)_1
\,(X^*X)_0^{-1/2}\,\partial_t(\g5{}X)_1\,(X^*X)_0^{-1/2}\Bigr)
\nonumber \\
&&\ \ +a^4\Tr\Bigl(\,\g5X_0\,(X^*X)_0^{-1/2}\,\partial_s(\g5{}X)_1
\,(X^*X)_2^{-1/2}\,\partial_t(\g5{}X)_1\,(X^*X)_0^{-1/2}\Bigr)
\nonumber \\
&&\ \ +a^4\Tr\Bigl(\,\g5X_0\,(X^*X)_0^{-1/2}\,\partial_s(\g5{}X)_1
\,(X^*X)_0^{-1/2}\,\partial_t(\g5{}X)_1\,(X^*X)_2^{-1/2}\Bigr)+O(a) 
\nonumber \\
&&\ =-\frac{1}{2}a^4(t^{cab}+t^{acb}+t^{abc})
\Tr\Bigl(\,L_0^{-5/2}\g5{}X_0\,\partial_s(\g5{}X)_1^a\,\partial_t(\g5{}X)_1^b
\,V_2^c\Bigr)+O(a)
\label{q20}
\ee
where we have set $t^{abc}=\tr(T^aT^bT^c)$. Modulo an $O(a)$ term, (\ref{q20})
is antisymmetric under interchange of $\partial_s(\g5{}X)_1^a$ and
$\partial_t(\g5{}X)_1^b$. To see this, note that $\partial_s(X^*X)_1^a=
\g5{}X_0\partial_s(\g5{}X)_1^a+\partial_s(\g5{}X)_1^a\g5{}X_0$.
Since $(X^*X)_1=L_1$ does not contain $\gamma^{\mu}$'s, it follows that
$\g5{}X_0\partial_s(\g5{}X)_1^a$ can be replaced by 
$-\partial_s(\g5{}X)_1^a\g5{}X_0$ in (\ref{q20}). The claimed antisymmetry
then follows from the cyclicity of the trace after using (v) above. 
Taking this into account in (\ref{x6}), and noting
$\frac{1}{2}d^{abc}=t^{cab}+t^{cba}=t^{acb}+t^{bca}=t^{abc}+t^{bac}\,$,
we get
\be
\Tr(P\lb\partial_sP,\partial_tP\rb)
=-\frac{3}{32}a^4d^{abc}\Tr\Bigl(\,L_0^{-5/2}\g5{}X_0\,\partial_s(\g5{}X)_1^a
\,\partial_t(\g5{}X)_1^b\,V_2^c\Bigr)+O(a)
\label{q21}
\ee
We calculate
\be
\partial_s(\g5{}X)_1^a\,\partial_t(\g5{}X)_1^b
&=&\Bigl(\,-\gamma^{\mu}(\partial_s\nabla_{\mu})_1^a
+\frac{1}{2}r(\partial_s\Delta)_1^a\Bigr)
\Bigl(\gamma^{\nu}(\partial_t\nabla_{\nu})_1^b
+\frac{1}{2}r(\partial_t\Delta)_1^b\Bigr) \nonumber \\
&=&\wt{V}^{ab}\ +\ \mbox{a term not involving $\gamma^{\mu}$'s}
\label{q22}
\ee
where
\be
\wt{V}^{ab}&=&\wt{V}^{(1)ab}+\wt{V}^{(2)ab}
\label{q23} \\
\wt{V}^{(1)ab}&=&-\frac{1}{2}r\gamma^{\mu}
\Bigl(\,(\partial_s\nabla_{\mu})_1^a(\partial_t\Delta)_1^b
-(\partial_s\Delta)_1^a(\partial_t\nabla_{\mu})_1^b\Bigr)
\label{q24} \\
\wt{V}^{(2)ab}&=&-\frac{1}{2}\lb\gamma^{\mu},\gamma^{\nu}\rb
(\partial_s\nabla_{\mu})_1^a(\partial_t\nabla_{\nu})_1^b
\label{q25}
\ee
It follows from (\ref{q21}) and (\ref{q22})--(\ref{q25}) that
\be
&&\Tr(P\lb\partial_sP,\partial_tP\rb)
=\frac{-3}{32}a^4d^{abc}\Tr(L_0^{-5/2}\g5{}X_0\,\wt{V}^{ab}\,V_2^c)+O(a)
\nonumber \\
&&\quad=\frac{-3}{32}a^4d^{abc}\Tr\Bigl(\,L_0^{-5/2}\g5
\Bigl(\,\gamma^{\mu}\nabla_{\mu}(\wt{V}^{(1)ab}V_2^{(2)c}
+\wt{V}^{(2)ab}V_2^{(1)c})+r\frac{1}{2}\Delta\wt{V}^{(2)ab}V_2^{(2)c}\Bigr)
\Bigr) \nonumber \\
&&\qquad+O(a)
\label{q26}
\ee
$V^{(1)ab}$ and $V^{(2)ab}$ can be determined as follows.
Recalling $\partial_sU_{\mu}(x)=a\partial_sA_{\mu}(x)+O(a^2)$ we get
\be
(\partial_s\nabla_{\mu}^{\pm})_1\psi(x)=(\partial_sA_{\mu}(x)+O(a))
\psi(x\pm{}ae_{\mu})
\label{q27}
\ee
and calculations give
\be
(\partial_s\nabla_{\mu})_1^a(\partial_t\nabla_{\nu})_1^b\psi(x)
&=&\frac{1}{4}(\partial_sA_{\mu}^a(x)\partial_tA_{\nu}^b(x)+O(a))
\nonumber \\
&&\ \times\bigg\lb{\psi(x+ae_{\mu}+ae_{\nu})+\psi(x+ae_{\mu}-ae_{\nu})
\atop +\psi(x-ae_{\mu}+ae_{\nu})+\psi(x-ae_{\mu}-ae_{\nu})}\bigg\rb
\label{q28} 
\ee
\be
\Bigl(\,(\partial_s\nabla_{\mu})_1^a(\partial_t\Delta)_1^b
\!\!&-&\!\!(\partial_s\Delta)_1^a(\partial_t\nabla_{\mu})_1^b\Bigr)\psi(x)
\nonumber \\
&=&-\sum_{\nu}\frac{1}{2}
\Bigl(\partial_sA_{\mu}^a(x)\partial_tA_{\nu}^b(x)
-\partial_sA_{\nu}^a(x)\partial_tA_{\mu}^b(x)+O(a)\Bigr) \nonumber \\
&&\quad\times
\bigg\lb{\psi(x+ae_{\mu}+ae_{\nu})+\psi(x+ae_{\mu}-ae_{\nu})
\atop +\psi(x-ae_{\mu}+ae_{\nu})+\psi(x-ae_{\mu}-ae_{\nu})}\bigg\rb
\label{q29}
\ee
which determine $\wt{V}^{(1)ab}$ and $\wt{V}^{(2)ab}$ in 
(\ref{q24})--(\ref{q25}). On the other hand, from (\ref{x10})--(\ref{x11})
and (\ref{q7})--(\ref{q10}) we find that $V_2^{(1)c}$ and $V_2^{(2)c}$
coincide up to $O(a)$ with $\wt{V}^{(1)ab}$ and $\wt{V}^{(2)ab}\,$,
respectively, after replacing $\partial_sA_{\mu}^a(x)\partial_tA_{\nu}^b(x)$
with $-\frac{1}{2}F_{\mu\nu}^c(x)$.

Having determined the operators in (\ref{q26}) we can now evaluate the trace
by first tracing over spinor indices and then evaluating the remaining
trace in the plane wave orthonormal basis $\{\phi_k(x)\}$ for periodic
scalar fields, given by
\be
&&\phi_k(x)=\frac{1}{\N}\,e^{ik\cdot{}x}
\label{q30} \\
&\N&=(2N)^4\qquad\quad,\qquad{}k_{\mu}\in\frac{\pi}{aN}
\{-N,-N+1,\dots,N-1\}
\label{q31}
\ee
The result is 
\be
\Tr(P\lb\partial_sP,\partial_tP\rb)
=\frac{\I}{32\pi^2}d^{abc}a^4\sum_x\e_{\mu\nu\rho\sigma}
\partial_sA_{\mu}^a(x)\partial_tA_{\nu}^b(x)F_{\rho\sigma}^c(x)
\label{q32}
\ee
where
\be
\I&=&\sum_ka^4\,\Delta^4k\,\I_r(ak) \label{q33} \\
&&\Delta^4k\;\equiv\;\frac{(2\pi)^4}{a^4\N}
=\mbox{the ``volume per $k$'' in (\ref{q31})}\,,\nonumber \\
&&\I_r(k)=\frac{-3r}{8\pi^2}\,
\frac{\prod_{\nu=1}^4\cos{}k_{\nu}\Big\lb\,-1+\sum_{\mu}(1-\cos{}k_{\mu})
-\sum_{\mu}\frac{\sin^2k_{\mu}}{\cos{}k_{\mu}}\Big\rb}
{\Big\lb\sum_{\mu}\sin^2k_{\mu}+r^2\Bigl(\,-1+\sum_{\mu}(1-\cos{}k_{\mu})
\Bigr)^2\,\Big\rb^{5/2}}
\label{q34}
\ee
(The denominator in this expression comes from $L_0^{-5/2}$ in 
(\ref{q26}); we have used (\ref{q11}).)
Changing variables from $k$ to $ak$ in (\ref{q33}) leads to
\be
\lim_{a\to0}\;\I=\int_{-\pi}^{\pi}d^4k\,\I_r(k)\,.
\label{q35}
\ee
This integral was encountered in \cite{DA-Suzuki} in connection with the
axial anomaly for fermions with Overlap Dirac operator, and was found
to equal 1 (independent of $r$). We can now take the $a\to0$ limit
in (\ref{q32}) and get the desired result:
\be
\lim_{a\to0}\;\Tr(P\lb\partial_sP,\partial_tP\rb)
=\frac{-1}{32\pi^2}d^{abc}\int_{T^4}d^4x\,\e_{\mu\nu\rho\sigma}
\partial_sA_{\mu}^a(x)\partial_tA_{\nu}^b(x)F_{\rho\sigma}^c(x)
\label{q37}
\ee
In deriving this formula we have followed the approach used in 
\cite{DA-Suzuki} for calculating the axial anomaly for the overlap Dirac
operator. Presumably the other approaches of ref.'s 
\cite{KY} and \cite{Fuji} could also be used to derive this formula.

\noindent {\it Remark.} The preceding also shows that
\be
\lim_{a\to0}\ \frac{1}{8}\tr\{\lb\e\lb\partial_s\e,\partial_t\e\rb\rb(x,x)\}
&=&\frac{-1}{32\pi^2}d^{abc}\e_{\mu\nu\rho\sigma}
\partial_sA_{\mu}^a(x)\partial_tA_{\nu}^b(x)F_{\rho\sigma}^c(x)
=\frac{-1}{24}\tr(\F(x,s,t))^3 \nonumber \\
&&\label{q38}
\ee
Here $\lb\cdots\rb(x,y)$ denotes the kernel of the
operator $\lb\cdots\rb\,$,
$\F(x,s,t)$ is the curvature of the gauge field 
$\A(x,s,t)=A_{\mu}^{(s,t)}(x)dx^{\mu}$ in 6 dimensions, and the last
coefficient arises as 
$2\pi{}i(\frac{1}{(2\pi{}i)^33!})=\frac{-1}{24\pi^2}$.
This result for the ``topological field'' 
$q(x,s,t)=\frac{1}{8}\tr\{\lb\e\lb\partial_s\e,\partial_t\e\rb\rb(x,x)\}$
in 6 dimensions was used in \cite{L2}\footnote{
In \cite{L2} a more general
$q(x,s,t)$ was considered; however, to obtain the mentioned result it
suffices to consider the present $q(x,s,t)$.} 
to show the
existence of a $w_+(U)$ (or equivalently, the existence of a local
gauge-invariant current $j_{\mu}(x)$ satisfying certain conditions) 
to all orders in an expansion in the lattice spacing
$a$ such that the corresponding overlap $\la{}v_+\,,w_+\ra$ is 
gauge-invariant when $d_R^{abc}=0$.
The calculations in this section leading to (\ref{q38}) above could
therefore be useful as a starting point for finding
explicit expressions for the terms in $j_{\mu}(x)$.

\section{Concluding remarks}

The main result of this paper is that the overlap formulation of chiral 
gauge theory on the lattice reproduces the global obstructions to 
gauge-invariance discussed in the continuum by Alvarez-Gaum\'e and Ginsparg 
\cite{AG}. We showed that the obstruction on the lattice reduces to the
continuum obstruction in the classical continuum limit. This, together
with the fact that the lattice obstruction is also an integer (winding
number), implies that the lattice obstruction coincides exactly
with the continuum one for small non-zero lattice spacing (i.e. close to
the classical continuum limit). Thus the overlap formulation is seen to
exactly capture topological structure of the continuum theory in
the nonabelian case, just as it does in the abelian case considered previously 
in \cite{Neu(PRD)}. We mention again that, while we have taken the spacetime
to be the 4-dimensional, our arguments and results generalise 
straightforwardly to Euclidean spacetime $T^{2n}$ for arbitrary $n$.

It might be instructive to
compare this with the situation for chiral Wilson fermions on the
lattice (where gauge-invariance is explicitly broken due to the Wilson
term, and only restored in the $a\to0$ limit). In this case the consistent
local anomaly 
$\frac{d}{d\theta}\log\det(D_{Wilson}^{\phi_{\theta}\cdot{}U})_+$ has been
shown to converge to the continuum anomaly in the classical continuum
limit \cite{anomaly(Wilson)}, so the integral 
\be
W_{Wilson}(\Phi)=\frac{1}{2\pi{}i}\int_0^1d\theta\,\frac{d}{d\theta}
\log\det(D_{Wilson}^{\phi_{\theta}\cdot{}U})_+ \nonumber
\ee
converges to the continuum
global obstruction $\deg(\Phi)$ in this limit. However, for non-zero
lattice spacing the integral $W_{Wilson}(\Phi)$ is {\it non-integer}
in general --it does not have a winding number interpretation since
$|\det(D_{Wilson}^U)_+|$ is not gauge-invariant. This is in contrast
to the overlap case where 
\be
W(\Phi)=\frac{1}{2\pi{}i}\int_0^1d\theta\,\frac{d}{d\theta}
\log\la{}v_+\,,w_+(\phi_{\theta}\cdot{}U)\ra \nonumber
\ee
is an integer (winding number) since $|\la{}v_+\,,w_+(U)\ra|$ is 
gauge-invariant and $\la{}v_+\,,w_+(\phi_{\theta}\cdot{}U)\ra$
differs from $\la{}v\,,w_+(U)\ra$ only by a phase factor. Thus, in contrast
to the overlap case, the non-integer $W_{Wilson}(\Phi)$ is in general not
equal to the integer-valued continuum obstruction $\deg(\Phi)$ at any 
non-zero lattice spacing; equality only occurs in the limit $a\to0$.

As a consequence of our main result, we found that the continuum anomaly
cancellation condition $d_R^{abc}=0$ is a necessary condition for anomaly
cancellation in the overlap on the lattice (at least when $Map(T^5,SU(N))$
contains maps with non-trivial degree). While this is no surprise,
our derivation is robust compared to other approaches: Firstly, 
it is independent of the choice of phase in the overlap (in contrast to
the consistent local anomaly which does depend on the phase choice), and
secondly, no approximations, or assumptions on the gauge field,
have been used.

In the continuum argument of ref. \cite{AG}, the winding number obstruction
$W(\Phi)$ is shown to equal the index of a Dirac operator $\D$ in 
$6\!=\!2n\!+\!2$ dimensions. The index theorem is then used to get
$\index\D=\deg{\Phi}$. In this paper we have followed a different route:
a lattice version of the determinant line bundle
approach of ref. \cite{AS-BF}. 
In fact, it is also possible to give a lattice version of the original 
argument of Alvarez-Gaum\'e and Ginsparg in the overlap setting, using a 
certain lattice
Dirac operator in $2n\!+\!2$ dimensions (with the extra 2 dimensions 
being continuous)
\cite{DA(top)}\footnote{In this case, the index density of the lattice $\D$
turns out to be the aforementioned topological field 
$q(x,s,t)$ in $6\!=\!2n\!+\!2$ dimensions
which appeared in ref. \cite{L2}.}

The obstructions of Alvarez-Gaum\'e and Ginsparg are but one type of 
obstruction to gauge-invariance of the chiral determinant. In general,
the obstructions are manifestations of non-trivial topological structure
of the determinant line bundle over the orbit space of gauge fields.
This topic has been studied in detail in the continuum; see, e.g.,
\cite{AS-BF}. The results of \cite{Neu(PRD)} and the present paper suggest
that in general the continuum topological structure of the determinant
line bundle can be reproduced on the lattice in the overlap formulation
(at least when the spacetime manifold is a $2n$-dimensional torus).
The determinant line bundle comes equipped with a canonical $U(1)$
connection, and the difference $Im\Gamma(A^{(1)})-Im\Gamma(A^{(0)})$
for the effective action 
$\Gamma(A)=\log\,\det(D_+^A)$ can be expressed in
terms of the parallel transport of this connection along a path joining
$A^{(0)}$ to $A^{(1)}$ \cite{AG-DP}.
This can in turn be expressed in terms of a
spectral flow ($\eta$-invariant) of a Dirac operator and a Chern-Simons
term, both in $5\!=\!2n\!+\!1$ dimensions. 
Lattice versions of these relations in 
the overlap setting have already been found \cite{Suzuki(rel),Kiku}.

Finally, it is interesting to note that the quantity $\Tr(PdPdP)=
\frac{1}{8}\Tr(\e{}d\e{}d\e)\,$, which in the present setting appears as the
curvature of the overlap determinant line bundle 
(or the 'Berry curvature' in the terminology of ref. \cite{Neu(PRD)}),
also arises as the curvature of a determinant line bundle 
in canonical quantisation of the continuum theory.\footnote{
I thank Prof. J. Mickelsson for pointing this out to me.}
See \cite{Mick} and the ref.'s therein. In that setting one considers
a certain infinite-dimensional Grassmannian manifold consisting of 
splittings $V_+\oplus{}V_-$ of the Hilbert space of 1-particle states; 
each splitting corresponds to an $\e=P_{V_+}-P_{V_-}$. 
There is a canonical determinant
line bundle on this manifold, and its curvature turns out to be a 
renormalised version of $\frac{1}{8}\Tr(\e{}d\e{}d\e)$.\footnote{ 
In fact, a lattice regularisation of chiral gauge theory
(different to the overlap) in which this
quantity also appears has been presented in ref. \cite{Mick(lattice)}.
Gauge-invariance appears to be problematic in this approach though: 
the notion of gauge
symmetry needs to be modified on the lattice in a way that involves
non-local operators.}
It could be interesting to explore the apparent analogy between
this continuum formulation and the lattice overlap 
formulation.
Recently, an obstruction to canonical quantisation of the continuum
theory on odd-dimensional spacetimes was described in \cite{Carey}.
Instead of the 'Berry curvature' 2-cocycle,
the obstruction there is given in terms of a 3-cocycle known as the
Dixmier--Douadly class. It could be interesting to see if there is something
analogous to this in the lattice overlap formulation.

\noindent {\it Acknowledgements.} This work was presented at the 
mini-workshop ``New developments in lattice gauge theory'', CSSM, Adelaide,
4-5 April'00, and I thank Prof.'s 
Ting-Wai Chiu, Kazuo Fujikawa, Urs Heller, Jouko Mickelsson and 
Tony Williams for interesting comments and discussions at that time.
I also thank Prof. Herbert Neuberger for correspondence on the Overlap
and, together with Prof.'s Martin L\"uscher and Jun Nishimura,
for feedback on earlier versions of this paper.
The author is supported by an ARC postdoctoral fellowship.

\appendix
\section*{Appendix}
\section{Derivation of (\ref{a7}):
$\frac{\la{}v_+\,,(dw_+)_{\perp}\ra}{\la{}v_+\,,w_+\ra}=\Tr(dDD^{-1}P_+)\,$.}

Let $U(t)$ be a smooth curve in $\U$. Using (\ref{2.14}) we calculate
\be
\frac{d}{dt}\,\log\,\la{}v_+\,,w_+\ra
&=&\frac{d}{dt}\,\log\,\la{}v_+\,,\wh{D}w_+\ra \nonumber \\
&=&\frac{1}{\la{}v_+\,,\wh{D}w_+\ra}\Bigl(\,
\la{}v_+\,,{\textstyle \frac{d}{dt}}\wh{D}w_+\ra+
\la{}v_+\,,\wh{D}{\textstyle \frac{d}{dt}}w_+\ra\Bigr)
\label{A1}
\ee
At $t=0$ (and with our assumption that $D$ has no zero-modes) we have
\be
\la{}v_+\,,{\textstyle \frac{d}{dt}}\wh{D}w_+\ra
&=&\la{}v_+\,,{\textstyle \frac{d}{dt}}\wh{D}\,\wh{D}^{-1}\wh{D}w_+\ra
=\la{}v_+\,,\wh{D}w_+\ra\,
\la{}v_+\,,{\textstyle \frac{d}{dt}}\wh{D}\,\wh{D}^{-1}v_+\ra
\nonumber \\
&=&\la{}v_+\,,\wh{D}w_+\ra\,\frac{d}{dt}\Big|_{t=0}
\det\Bigl(\,D(t)D(0)^{-1}\Big|_{\C_+}\Bigr) \nonumber \\
&=&\la{}v_+\,,\wh{D}w_+\ra\,\Tr({\textstyle \frac{d}{dt}}DD^{-1}P_+)
\nonumber
\ee
so the first term on the right-hand side of (\ref{A1}) is
$\Tr(\frac{d}{dt}DD^{-1}P_+)$. Comparing (\ref{A1}) with (\ref{a6}) we
see that (\ref{a7}) holds iff
\be
\la{}v_+\,,\wh{D}{\textstyle \frac{d}{dt}}w_+\ra
=\la{}v_+\,,\wh{D}w_+\ra\,\la{}w_+\,,{\textstyle \frac{d}{dt}}w_+\ra
\label{A2}
\ee
Choose an orthonormal basis $w_1(t),\dots,w_d(t)$ for $\C_+^{(1,U(t))}$
such that $w_+=w_1\wedge\cdots\wedge{}w_d$. Then
\be
\la{}v_+\,,\wh{D}{\textstyle \frac{d}{dt}}w_+\ra
=\sum_{j=1}^d\la{}v_+\,,\wh{D}(w_1\wedge\cdots\wedge{\textstyle \frac{d}{dt}}
w_j\wedge\cdots\wedge{}w_d)\ra
\label{A3}
\ee
Substitute $\frac{d}{dt}w_j=\sum_{k=1}^d\la{}w_k\,,\frac{d}{dt}w_j\ra{}w_k
+(\frac{d}{dt}w_j)_{\perp}$ in (\ref{A3}), where 
$(\frac{d}{dt}w_j)_{\perp}\in\C_-^{(1,U(t))}$.
Since $D$ maps $\wh{\C}_{\pm}=\C_{\mp}^{(1,U)}$ to $\C_{\mp}\,$, the terms 
involving $(\frac{d}{dt}w_j)_{\perp}$ give vanishing contribution, and we get
\be
\la{}v_+\,,\wh{D}{\textstyle \frac{d}{dt}}w_+\ra
&=&\sum_{j=1}^d\la{}v_+\,,\wh{D}w_+\ra\,\la{}w_j\,,{\textstyle \frac{d}{dt}}
w_j\ra\,.
\nonumber
\ee
This equals the right-hand side of (\ref{A2}) as required, since
\be
\la{}w_+\,,{\textstyle \frac{d}{dt}}w_+\ra
&=&\sum_{j=1}^d\la{}w_+\,,w_1\wedge\cdots\wedge{\textstyle \frac{d}{dt}}w_j
\wedge\cdots\wedge{}w_d\ra \nonumber \\
&=&\sum_{j=1}^d\la{}w_+\,,w_+\ra\,\la{}w_j\,,{\textstyle \frac{d}{dt}}w_j\ra
=\sum_{j=1}^d\la{}w_j\,,{\textstyle \frac{d}{dt}}w_j\ra\,.
\nonumber
\ee

\section{Derivation of (\ref{a13})} 

\noindent It suffices to restrict to a surface
in $\U$ with coordinates $(s,t)$ and show
\be
\la{}\partial_sw_+,\partial_tw_+\ra-\la{}\partial_tw_+,\partial_sw_+\ra
=\Tr(P\partial_sP\partial_tP)-\Tr(P\partial_tP\partial_sP)
\label{B1}
\ee
Let $w_1(s,t),\dots,w_d(s,t)$ be an orthonormal basis for $\C_+^{(1,U(s,t))}$
such that $w_+=w_1\wedge\cdots\wedge{}w_d$. It is convenient to use
bra-ket notation:
\be
P=\sum_{k=1}^d|w_k\ra\la{}w_k|\qquad\quad,\qquad\
\partial_sP=\sum_{k=1}^d|\partial_sw_k\ra\la{}w_k|+|w_k\ra\la\partial_sw_k|
\nonumber
\ee
Then
\be
\Tr(P\partial_sP\partial_tP)
&=&\sum_{j=1}^d\la{}w_j|\partial_sP\partial_tP|w_j\ra \nonumber \\
&=&\sum_{j,k,l=1}^d\la{}w_j|\Bigl(\,|\partial_sw_k\ra\la{}w_k|
+|w_k\ra\la\partial_sw_k|\Bigr)
\Bigl(\,|\partial_tw_l\ra\la{}w_l|+|w_l\ra\la\partial_tw_l|\Bigr)|w_j\ra
\nonumber \\
&=&\sum_{j,k}\la{}w_j|\partial_sw_k\ra\la{}w_k|\partial_tw_j\ra+
\sum_{j,k}\la{}w_j|\partial_sw_k\ra\la\partial_tw_k|w_j\ra
\nonumber \\
& &\ +\sum_j\la\partial_sw_j|\partial_tw_j\ra+
\sum_{k,l}\la\partial_sw_j|w_l\ra\la\partial_tw_l|w_j\ra
\label{B2}
\ee
The first and fourth sums are clearly symmetric under 
$\partial_s\leftrightarrow\partial_t$.
The second sum is likewise symmetric under 
$\partial_s\leftrightarrow\partial_t$ as is easily
seen using $\la\delta{}w_j|w_k\ra+\la{}w_j|\delta{}w_k\ra=
\delta\la{}w_j|w_k\ra=0$. It follows that
\be
\Tr(P\partial_sP\partial_tP)-\Tr(P\partial_tP\partial_sP)
=\sum_j\la\partial_sw_j|\partial_tw_j\ra-\la\partial_tw_j|\partial_sw_j\ra
\nonumber
\ee
and this is equal to the left-hand side of (\ref{B1}) as required.

\end{document}